\documentclass[12pt]{article}
\newcommand{\pbv}{\,^{+}\!\mbox{\boldmath$v$}}
\newcommand{\mbv}{\,^{-}\!\mbox{\boldmath$v$}}
\newcommand{\pmbpi}{\,^{\pm}\!\mbox{\boldmath$\pi$}}
\newcommand{\pmbtau}{\,^{\pm}\!\mbox{\boldmath$\tau$}}
\newcommand{\bc}{\mbox{\boldmath$c$}}
\newcommand{\bpi}{\mbox{\boldmath$\pi$}}
\newcommand{\btau}{\mbox{\boldmath$\tau$}}
\newcommand{\bv}{\mbox{\boldmath$v$}}

\begin{document}

\title{On the length expectation values in quantum Regge
calculus}
\author{V.M.Khatsymovsky \\
 {\em Budker Institute of Nuclear Physics} \\ {\em
 Novosibirsk,
 630090,
 Russia}
\\ {\em E-mail address: khatsym@inp.nsk.su}}
\date{}
\maketitle
\begin{abstract}
Regge calculus configuration superspace can be embedded
into a more general superspace where the length of any edge
is defined ambiguously depending on the 4-tetrahedron
containing the edge. Moreover, the latter superspace can be
extended further so that even edge lengths in each the
4-tetrahedron are not defined, only area tensors of the
2-faces in it are. We make use of our previous result
concerning quantisation of the area tensor Regge calculus
which gives finite expectation values for areas. Also our
result is used showing that quantum measure in the Regge
calculus can be uniquely fixed once we know quantum measure
on (the space of the functionals on) the superspace of the
theory with ambiguously defined edge lengths. We find that
in this framework quantisation of the usual Regge calculus
is defined up to a parameter. The theory may possess
nonzero (of the order of Plank scale) or zero length
expectation values depending on whether this parameter is
larger or smaller than a certain value. Vanishing length
expectation values means that the theory is becoming
continuous, here {\it dynamically} in the originally
discrete framework.
\end{abstract}

\newpage
In our previous works \cite{Kha,Kha1} we have developed the
viewpoint on quantisation of the Regge calculus as a
particular case of a more general system, the area tensor
Regge calculus. In \cite{Kha} the area tensor Regge
calculus has been quantised and shown to lead to finite
expectation values of areas. In \cite{Kha1} it has been
shown that quantisation of the ordinary Regge calculus is
uniquely defined under natural physical assumptions
provided that this system is considered as a particular
case of the system where the same edge can have different
lengths in the different 4-tetrahedra containing it.
Quantisation of the latter system is considered to be
already defined in the form of a quantum measure.

Now we would like to combine these results to define
quantum measure in the Regge calculus proceeding from the
area tensor Regge calculus whose quantisation (issued from
the canonical approach) is consistently definable. To do
this yet one step is required since we must relate area
tensor Regge calculus and the above mentioned
generalisation of Regge calculus with multiply defined edge
lengths. Equivalently to say, the latter system can be
viewed as collection of the 4-simplices not necessarily
having the same edge lengths on junctions between them; at
the same time, a 4-simplex in area tensor Regge calculus
does not even possess certain edge lengths or metric, only
tensors of the 2-faces.

We begin with the result of \cite{Kha} according to which
the vacuum expectation value of a functional on the set of
area tensors $\pi$ and connection matrices $\Omega$ in the
Euclidean signature case takes the form
\begin{eqnarray}
\label{VEV}
<\Psi (\{\pi\},\{\Omega\})> & = & \int{\Psi (-i\{\pi\},
\{\Omega\})\exp{\left (-\!
\sum_{\stackrel{t-{\rm like}}{\sigma^2}}{\tau
_{\sigma^2}\circ
R_{\sigma^2}(\Omega)}\right )}}\nonumber\\
 & & \hspace{-20mm} \exp{\left (i
\!\sum_{\stackrel{\stackrel{\rm not}{t-{\rm like}}}{\sigma^2}}
{\pi_{\sigma^2}\circ
R_{\sigma^2}(\Omega)}\right )}\prod_{\stackrel{\stackrel{\rm
 not}{t-{\rm like}}}{\sigma^2}}{\rm d}^6
\pi_{\sigma^2}\prod_{\sigma^3}{{\cal D}\Omega_{\sigma^3}}
\nonumber\\
& \equiv & \int{\Psi (-i\{\pi\},\{\Omega\}){\rm d}
\mu_{\rm area}(-i\{\pi\},\{\Omega\})}.                   
\end{eqnarray}

\noindent Here $A\circ B$ $\stackrel{\rm def}{=}$ ${1\over
2}A^{ab}B_{ab}$. The field variables are the area tensors
$\pi_{\sigma^2}$ on the triangles $\sigma^2$ (the
2-simplices $\sigma^2$) and SO(4) connection matrices
$\Omega_{\sigma^3}$ on the tetrahedrons $\sigma^3$ (the
3-simplices $\sigma^3$). This looks as the usual
field-theoretical path integral expression with exception
of the following three points. First, occurence of the Haar
measure on the group SO(4) of connections on separate
3-faces ${\cal D}\Omega_{\sigma^3}$. This is connected with
the specific form of the kinetic term $\pi_{\sigma^2}\circ
{\Omega}^{\dag}_{\sigma^2}\dot{\Omega}_{\sigma^2}$ which
appears when one passes to the continuous limit along any
of the coordinate direction chosen as time. Here SO(4)
rotation $\Omega_{\sigma^2}$ serves to parameterise
limiting form of $\Omega_{\sigma^3}$ when $\sigma^3$ tends
to $\sigma^2$. Such form of the kinetic term in the general
3D discrete gravity model has been deduced by Waelbroeck
\cite{Wae}.

Second, there are the terms $\pi\circ R$ in the exponential
instead of the Regge action which in the exact connection
representation would be the sum of the terms with 'arcsin'
of the type $|\pi |\arcsin{(\pi\circ R/|\pi |)}$
\cite{Kha2}. The reason is that the exponential should be
the sum of the constraints times Lagrange multipliers in
order to fit the canonical quantisation prescription in the
continuous time limit whatever coordinate is taken as a
time. These constraints just do not contain 'arcsin' (in
empty spacetime; situation is more complicated in the
presence of matter fields!).

Third, occurence of a set of the triangles integrations
over area tensors $\tau_{\sigma^2}$ of which are absent.
It is the set of those triangles the curvature matrices on
which are functions, via Bianchi identities, of all the
rest curvatures. (Integrations over them might result in
the singularities of the type of $[\delta(R-\bar{R})]^2$).
In the situations when one defines direction of the
coordinate called "time", the $t$-like triangles turn out
to be possible choice of the above set (therefore
definition in terms of the Bianchi identities can probably
serve as definition of what set of triangles can be chosen
as timelike in general case).

Here we imply the situation allowing intuitively evident
definition of time direction, namely, a certain regular
structure \cite{MisThoWhe} of the Regge manifold considered
as that consisting of a sequence of the 3D Regge manifolds
$t$ = $const$ of the same structure (of linking different
vertices) usually called {\it the leaves of the foliation}
along $t$. The vertices of the considered 3D leaf will be
denoted as $i$, $k$, $l$, \dots. The $i^+$ means image in
the next-in-$t$ leaf of the vertex $i$ taken at the current
moment $t$. The notation $(A_1A_2\ldots A_{n+1})$ means
unordered $n$-simplex with vertices $A_1$, $A_2$, \ldots,
$A_{n+1}$ (triangle at $n$ = 2). The 4-simplices are
arranged into the 4-prisms between the
neighbouring 3D leaves. Let the 4-prism with bases $(iklm)$
and $(i^+k^+l^+m^+)$ consists of the 4-simplices
$(ii^+klm)$, $(i^+kk^+lm)$, $(i^+k^+ll^+m)$,
$(i^+k^+l^+mm^+)$.

With these definitions, we can divide the whole set of
$n$-simplices into the three groups, namely, first, those
of the type $(ii^+A_1\ldots A_{n-1})$ containing the edge
$(ii^+)$; second, those of the type $(i_1i_2\ldots
i_{n+1})$ completely located in the leaf; third, those
differing from the second type by occurence of the
superscript '+' on some of the vertices $i_1$, $i_2$,
\dots, $i_{n+1}$, e. g. the tetrahedron $(i^+klm)$. We
shall refer to these as to the $t$-{\it like}, {\it leaf}
and {\it diagonal} simplices, respectively. These terms can
be considered as the different values of a Regge analog of
the {\it world} index in general relativity (GR). The more
usual terms {\it timelike} and {\it spacelike} will be
reserved for the {\it local frame} indices $a$, $b$, $c$,
\dots of the tensors.

Let $v$ be tensor of any triangle, $\pi$ or $\tau$ ($\pi$
and $\tau$ are thus Regge analogs of spacelike and timelike
components of $v$ w. r. t. the world index).

Note that $v_{\sigma^2}$ means tensor taken in the local
frame of some $\sigma^4$ containing $\sigma^2$. For this
reason, the more detailed notation could be, e. g.,
$v_{\sigma^2|\sigma^4}$. If simplices are explicitly
defined as collections of vertices, we'll enumerate
vertices of $\sigma^4$ in the subscript and single out
among them by brackets the vertices corresponding to the
considered object; e. g. $v_{(ABC)DE}$ is tensor of the
triangle $(ABC)$ defined in the frame of the 4-simplex
$(ABCDE)$. Further, since the number of the triangles
$N_2$ is larger than the number of the 4-simplices $N_4$
generally there are more than one area tensors defined in
the frame of any given 4-simplex. If, say,
$v_{\sigma^2_1}$ and $v_{\sigma^2_2}$ are defined in the
same frame, we can form the scalar $v_{\sigma^2_1}\circ
v_{\sigma^2_2}$ which together with the ten areas of the
simplicial 2-faces leads to the additional conditions which
should be imposed in order that these tensors unambiguously
define ten simplicial edge lengths. That is, there are
points in the configuration space of the area tensor Regge
calculus which do not correspond to any flat 4-metric in a
given 4-simplex. On the other hand, in the simplicial
complex where the length of an edge is allowed to be
different in the different 4-simplices containing the edge
the area of a 2-face can be also different in the different
4-simplices containing the 2-face. At the same time, in our
formulation of area tensor Regge calculus each the 2-face
$\sigma^2$ possesses the area $|v_{\sigma^2}|$ $\equiv$
$(v_{\sigma^2}\circ v_{\sigma^2})^{1/2}$ as the only scalar
corresponding to it. That is, and vice versa, there are
also points in the configuration space of the Regge
calculus with independent simplicial edge lengths which do
not correspond to any point in the configuration space of
area tensor Regge calculus. Now, if we would like that a
configuration space of one theory be in correspondence with
a subset of the configuration space of another one, the
most natural would be to extend the configuration space of
area tensor Regge calculus by introducing into
consideration for each the 2-face $\sigma^2$ the tensors
$v_{\sigma^2|\sigma^4}$ for {\it all} the 4-simplices
$\sigma^4$ containing $\sigma^2$.

For the measure, such the extension looks trivial, as
simply adding, first, integrations
over ${\rm d}^6\pi_{\sigma^2|\sigma^4}$ other than
${\rm d}^6\pi_{\sigma^2}$. Being applied to the functionals
of $\pi_{\sigma^2}$ only, these new integrations result
simply in an infinite normalisation factor (some
intermediate regularisation which confines the limits of
integrals by some large although finite areas is implied).
Second, also integrations over ${\rm d}^6\tau_{\sigma^2|
\sigma^4}$ should be inserted. Otherwise, if $\tau_{
\sigma^2|\sigma^4}$ were treated as parameters, knowing
these parameters would allow to almost completely fix
geometry of 3D leaves, i. e. dynamics itself.

A priori nothing prevent us to integrate over
${\rm d}^6\tau_{\sigma^2|\sigma^4}$ in the whole range of
$\tau_{\sigma^2|\sigma^4}$ as independent variables. At the
same time, the result of \cite{Kha} concerning existence of
the well-defined area expectation values was obtained just
in the assumption of smallness of the tensors $\tau$ as
compared to 1 (Plankian unity). Therefore for technical
reasons we choose to restrict absolute values of the
tensors $\tau$. In what follows, the conditions which say
that system is the usual Regge manifold are to be imposed,
and in this framework it is sufficient to restrict some 4
scalars connected with $\tau$ per vertex. This corresponds
to fixing lapse-shift vector in the continuum GR, i. e. to
fixing gauge. Nondegenerate such anzats amounts, say, to
fixing $|\tau_1|^2$, $|\tau_2|^2$, $|\tau_3|^2$, $\tau_1
\circ \tau_2$ for the 4 tensors $\tau_i$ at each vertex
$\sigma^0$. That is, $\tau_i$ are certain 4 functions of
the 0-simplex $\sigma^0$, and a more detailed notations
$\tau_i(\sigma^0)$ $\equiv$
$\tau_{\sigma^2_i(\sigma^0)|\sigma^4(\sigma^0)}$ imply
choice of the 4-simplex $\sigma^4(\sigma^0)$ and the 3
triangles $\sigma^2_i(\sigma^0)$ meeting at the $t$-like
edge at the vertex $\sigma^0$ and spanning this 4-simplex.

In view of the above consideration, the measure in the
extended configuration space of area tensor Regge calculus
takes the form
\begin{eqnarray}
\label{area_ext}
{\rm d}\mu_{\rm area\, extended} & = & \exp{\left (-\!
\sum_{\stackrel{t-{\rm like}}{\sigma^2}}{\tau
_{\sigma^2}\circ R_{\sigma^2}(\Omega)}
+ i\!\sum_{\stackrel{\stackrel{\rm not}
{t-{\rm like}}}{\sigma^2}}{\pi_{\sigma^2}\circ
R_{\sigma^2}(\Omega)}\right )}\nonumber\\
 & & \prod_{\sigma^0}{\left (\delta (\tau_1(\sigma^0)\circ
\tau_2(\sigma^0) - \zeta\varepsilon_1\varepsilon_2)
\prod^3_{i=1}{\delta (\tau_i(\sigma^0)\circ\tau_i(\sigma^0)
- \varepsilon^2_i)}\right )}\nonumber\\
 & & \left (\prod_{\sigma^2}{\prod_{\sigma^4\supset
\sigma^2}{{\rm d}^6v_{\sigma^2|\sigma^4}}}\right )
\prod_{\sigma^3}{{\cal D}\Omega_{\sigma^3}}              
\end{eqnarray}

\noindent where, remind, $v$ means $\pi$ or $\tau$;
0 $<$ $\varepsilon_i$ $\ll$ 1 and -1 $<$ $\zeta$ $<$ 1 are
parameters.

Thus, we have the system, area tensor Regge calculus
quantisation of which is defined in the form of quantum
measure (\ref{area_ext}) and which containes Regge calculus
with independent lengths as a particular case. In turn, the
latter system containes the ordinary Regge calculus. As a
result, the latter corresponds to a hypersurface
$\Gamma_{\rm Regge}$ in the configuration space of area
tensor Regge calculus. The quantum measure can be viewed as
a linear functional $\mu_{\rm area\, extended}(\Psi)$ on
the space of the functionals $\Psi (\{\pi\})$ on the
configuration space (for our purposes it is sufficient to
restrict ourselves by dependence on the set of area tensors
$\{\pi\}$, especially as the connection matrices $\Omega$
can be more general quantities than simply rotations
connecting the neighbouring simplicial frames and their
physical interpretation is difficult). Physical assumption
is that we can consider ordinary Regge calculus as a kind
of a state of this more general area tensor system. We can
associate with this state the functional of the form
\begin{equation}
\Psi (\{\pi\}) = \psi (\{\pi\})\delta_{\rm Regge}(\{\pi\})
\end{equation}                                           

\noindent where $\delta_{\rm Regge}(\{\pi\})$ is
(many-dimensional) $\delta$-function with support on
$\Gamma_{\rm Regge}$. The derivatives of $\delta_{\rm
Regge}$ have the same support, but these would violate
positivity in what follows. To be more presize,
delta-function is not a function, but can be viewed as such
if regularised. If the measure on such the functionals
exists in the limit when this regularisation is removed,
this allows to define the quantum measure on $\Gamma_{\rm
Regge}$,
\begin{equation}
\mu_{\rm Regge}(\cdot) = \mu_{\rm area\, extended}(\delta
_{\rm Regge}(\{\pi\})~\cdot).                            
\end{equation}

The $\delta$-function $\delta_{\rm Regge}$ is equal to the
product of the two deltas,
\begin{equation}
\delta_{\rm Regge}(\{\pi\}) = \delta_{\rm cont}(\{\pi\})
\delta_{\rm metric}(\{\pi\}).                            
\end{equation}

\noindent Here $\delta_{\rm metric}(\{\pi\})$ singles out
hypersurface in the configuration space of the area tensor
Regge calculus corresponding to the system with
well-defined (although independent) metrics in the
different 4-simplices; the $\delta_{\rm cont}(\{\pi\})$
singles out hypersurface in the configuration space of the
Regge calculus with independent lengths (metrics)
corresponding to the usual Regge calculus.

The $\delta_{\rm cont}$ has been considered in our previous
work \cite{Kha1}, and now the question is about $\delta
_{\rm metric}$. Here situation is even simpler, because the
problem reduces to that for a one 4-simplex and $\delta
_{\rm metric}$ is the product over separate 4-simplices.
In each the 4-simplex we should write out the constraints
which enable area tensors to correspond to certain
simplicial metric. Introduce temporarily (locally, in a
given 4-simplex) a world index. Denote the vertices of a
4-simplex by 0, 1, 2, 3, 4 so that (04) is $t$-like edge.
Denote tensor of the triangle $(\lambda\mu 4)$ ($\lambda$,
$\mu$, $\nu$, \dots = 0, 1, 2, 3) as $v^{ab}_{\lambda\mu}$.
Then conditions which constrain area tensors to be {\it
bivectors} corresponding to certain tetrad of vectors
attributed to the four edges $(4\lambda )$ take the form
\begin{equation}
\label{tetrad}
\epsilon_{abcd}v^{ab}_{\lambda\mu}v^{cd}_{\nu\rho}
\sim\epsilon_{\lambda\mu\nu\rho}.                        
\end{equation}

In the 36-dimensional configuration minisuperspace of the 6
antisymmetric tensors\footnote{Also there are the linear
constraints of the type $\sum{\pm v}$ = 0 which enable
closeness of the 3-faces of our 4-simplex. These
constraints are supposed to be already resolved.} $v^{ab}
_{\lambda\mu}$ the 20 eqs. (\ref{tetrad}) define the
16-dimensional hypersurface $\gamma (\sigma^4)$. The
$\delta_{\rm metric}$ is the product of $\delta$-functions
with
support on $\gamma (\sigma^4)$ over all the 4-simplices
$\sigma^4$. The covariant form of the constraints
(\ref{tetrad}) w. r. t. the world index means that these
$\delta$-functions are scalar densities of certain weight
w. r. t. the world index. In the continuum limit, this
means that limiting measure also behaves as scalar density
at the diffeomorphisms. This meets usual requirements for
the continuum measure, but for the weight of this scalar
density different values are possible corresponding to the
different factors $(\det{\|g_{\lambda\mu}\|})^{\alpha}$ in
the measure \cite{Mis,DeW}. In Regge calculus, this
corresponds to inserting factors of the type of
$V_{\sigma^4}^{\eta}$ where $V_{\sigma^4}$ is the 4-volume
and $\eta$ is a parameter. Thus $\delta_{\rm metric}$ takes
the form
\begin{equation}
\label{delta-metric}
\delta_{\rm metric} = \prod_{\sigma^4}{\int{V_{\sigma^4}
^{\eta}\delta^{21}(\epsilon_{abcd}v^{ab}_{\lambda\mu |
\sigma^4}v^{cd}_{\nu\rho |\sigma^4} - V_{\sigma^4}\epsilon
_{\lambda\mu\nu\rho})\,{\rm d}V_{\sigma^4}}}.            
\end{equation}

We can also introduce into consideration the measures which
are the prototype of that one suggested by Leutwyler
\cite{Leu} and reproduced by Fradkin and Vilkovisky in the
corrected approach \cite{FraVil}, $(\det{\|g_{\lambda\mu}
\|})^{-{3/2}}g^{00}{\rm d}^{10}g_{\lambda\mu}$. For
that we should insert $V_{\sigma^3}^2$ into
(\ref{delta-metric}) {\it per vertex} where $\sigma^3$ is
some leaf 3-face at this vertex. This just leads to
$g^{00}\det{\|g_{\lambda\mu}\|}$ in the continuum limit.
However, there are more than one way to attribute 3-face in
the leaf to a given vertex. This makes choice of the total
factor in the measure highly ambiguous, and each such
choice would violate equivalence of the different
3-simplices. Thus, it is the measures of the type of
$(\det{\|g_{\lambda\mu}\|})^{\alpha}{\rm d}^{10}
g_{\lambda\mu}$ which have natural Regge analogs\footnote{
Note, however, that, as it is shown below, the situation
with backward passing from the constructed measure to the
continuum one is more complex than simply getting $(\det{\|
g_{\lambda\mu}\|})^{\alpha}{\rm d}^{10}g_{\lambda\mu}$ in
the continuum limit. Rather we obtain this limiting form
only if summation in the functional integral is performed
over Regge manifolds with a fixed regular structure; if all
structures are taken into account, we get different
effective exponentials $\alpha$ for the different
functionals averaged with the help of this measure.}. On
the other hand, the role of special noninvariant form of
the measure $(\det{\|g_{\lambda\mu}\|})^{-{3/2}}g^{00}{\rm
d}^{10}g_{\lambda\mu}$ was shown in \cite{FraVil} to amount
to cancellation of all the (UV) divergences of the type
$\delta^{(4)}(0)$ in the effective action which arise in
the theory due to it's nonlinearity. These terms arise as
coincidence limit of some bilocals. This situation is
specific for the continuum and does not take place in the
discrete framework; therefore it is reasonable to confine
ourselves to the simple scalar density form of the measure.

Above we have described general form of the quantum measure
in Regge calculus considered as hypersurface in the
superspace of a more general theory. For performing the
further estimate it is convenient to make self-antiselfdual
decomposition of area tensors, so that $v_{\sigma^2}$ maps
into 3-vectors $\pbv_{\sigma^2}$ and $\mbv_{\sigma^2}$.
Locally in the given 4-simplex we introduce for the tensors
of the leaf/diagonal $v_{\alpha\beta}$ $\equiv$
$\pi_{\alpha\beta}$ ($\alpha$, $\beta$, $\gamma$, \dots =
1, 2, 3) or $t$-like $v_{0\alpha}$ $\equiv$
$\tau_{0\alpha}$ triangles the notations
\begin{equation}
\epsilon_{\alpha\beta\gamma}\pmbpi^{\gamma}\stackrel{\rm
def}{=}\pmbpi_{\alpha\beta},~~~\pmbtau_{\alpha}\stackrel{
\rm def}{=}\pmbtau_{0\alpha}.
\end{equation}                                           

\noindent Then we have
\begin{equation}
\label{tau-pi}
\pmbtau_{\alpha}=\epsilon_{\alpha\beta\gamma}c^{\beta}
\pmbpi^{\gamma}+{1\over 2}C\epsilon_{\alpha\beta\gamma}
\pmbpi^{\beta}\times\pmbpi^{\gamma}                      
\end{equation}

\noindent as general solution to the constraints
(\ref{tetrad}). Choose for definiteness selfdual components
and suppress index '+'. For the integral $\int{(\cdot)
\delta_{\rm metric}\prod_{\sigma^4,\sigma^2\subset\sigma^4}
{{\rm d}^{36}v_{\sigma^2|\sigma^4}}}$ we find the product
over the 4-simplices of the factors of the type
\begin{equation}                                        
\label{int-delta-metric}
\int{(\cdot)C^{\eta-6}[\bpi^1\times\bpi^2\cdot\bpi^3]
^{\eta-6}{\rm d}C{\rm d}^3\!\bc\,{\rm d}^9\bpi{\cal DO}}
\end{equation}

\noindent where ${\cal O}$ is an SO(3) rotation which
connects selfdual and antiselfdual sectors.

Let us address now the question of convergence of the
functional integral and estimate typical area (or length)
expectation values. Denote the scale of tensors of the
leaf/diagonal triangles $\pi$ in the given $\sigma^4$ by
$x$. Taking into account existence of yet another scale
$\varepsilon$ introduced when fixing lapse-shift
($\varepsilon_i$ in (\ref{area_ext})) we see that $C$,
$\bc$ parameterising $\btau$ by means of the decomposition
(\ref{tau-pi}) have the scale $x^{-2}$ and $x^{-1}$,
respectively (more accurately, $C$ $\sim$ $\varepsilon
/x^2$ and $\bc$ $\sim$ $\varepsilon /x$). Integrations
(\ref{int-delta-metric}) can be reduced to that one over
$x^{\eta -3}{\rm d}x$ and a number of compact ones.

Further, by changing variables a part of the integrations
over ${\cal D}\Omega_{\sigma^3}$ can be converted to the
integrations over ${\cal D}R_{\sigma^2}$ with $\sigma^2$
running over all the leaf and diagonal 4-simplices. These
integrations, as considered in \cite{Kha}, can be
factorised at small $\btau$ (corresponding to the choice
$\varepsilon$ $\ll$ 1). The integrals obtained can be
further decomposed into those in the self- and antiselfdual
sectors and read
\begin{equation}                                        
\int{e^{\textstyle i\pi\circ R}{\cal D}R} = {\nu (|\bpi |)^
2\over |\bpi |^4},~~~~\nu (l)={2l \over \pi}\int\limits_{0}
^{\pi}{e^{\textstyle -l/\sin{\varphi}}\:{\rm d}\varphi}.
\end{equation}

\noindent Therefore, if tensor of the considered $\sigma^2$
is defined in the given $\sigma^4$, we have, in addition to
the above $x^{\eta -3}{\rm d}x$, the factor which behaves
as $x^{-2}e^{-\lambda x}$ where $\lambda$ is a positive
bounded from below function of some parameters integrations
over which are to be made. For the purposes of studying
convergence of expressions defining expectation values we
consider simple calculational model replacing $e^{-\lambda
x}$ by $e^{-x}$. Evidently, if more than one area tensor is
defined in the frame of $\sigma^4$, the corresponding power
of $x^{-2}e^{-x}$ should be taken into account as a factor.
Consider some regular way of assigning area tensors to the
local frames. If $N^{(d)}_k$ is the number of the
$k$-simplices in the $d$-dimensional Regge manifold then
$N^{(3)}_2$ = $2N^{(3)}_3$. This means that in the leaf we
can attribute to each 3-simplex some two of it's 2-faces.
For example, let $(ikl)$ and $(ikm)$ be attributed to
$(iklm)$. Define their area tensors in some one of the two
4-simplices meeting at $(iklm)$, say, in the "future" one;
let the latter appears to be, say, $(ii^+klm)$. Then we
write on some of the vertices $i$, $k$, $l$, $m$, \dots in
these relations the superscript '+', i. e. shift them to
the next time leaf. This defines in a regular way in what
4-simplex the area tensor of each 2-simplex is defined.
Namely, $(i^+kl)$, $(i^+km)$ are defined in $(i^+kk^+lm)$;
$(i^+k^+l)$, $(i^+k^+m)$ are defined in $(i^+k^+ll^+m)$; no
triangles are defined in $(i^+k^+l^+mm^+)$. These are the
simplices in the 4-prism with bases $(iklm)$ and
$(i^+k^+l^+m^+)$, and in other prisms the area tensors may
be chosen to be defined in the same manner, if periodic
structure is implied. Of course, the above notations imply
that the corresponding simplices exist indeed; a definite
one of 24 possible ways of division of the 4-prism into the
four 4-simplices is taken. If $x$ is the area scale in the
4-simplex, say, $(i^+kk^+lm)$ in the frame of which the two
area tensors are defined, the corresponding factor in the
measure for it is
\begin{equation}                                        
x^{\eta -3}{\rm d}x\cdot (e^{-x}x^{-2})^2 = e^{-2x}x^{\eta
-7}{\rm d}x.
\end{equation}

Next introduce into consideration $\delta_{\rm cont}$ and
integrations not performed thus far. The $\delta_{\rm
cont}$ has been found \cite{Kha1} to read symbolically
\begin{equation}                                        
\delta_{\rm cont}=\prod_{\sigma^3}{V^4_{\sigma^3}\delta^6
(\Delta_{\sigma^3}S_{\sigma^3})}\left (\prod_{\sigma^2}
{V^3_{\sigma^2}\delta^3(\Delta_{\sigma^2}S_{\sigma^2})}
\right )^{-1}\prod_{\sigma^1}{V^2_{\sigma^1}\delta
(\Delta_{\sigma^1}S_{\sigma^1})}.
\end{equation}

\noindent Here $S_{\sigma^k}$ is the edge component metric
\cite{PirWil} on the $k$-simplex $\sigma^k$ (simply
collection of ${\scriptstyle k}{k+1\over 2}$ edge lengths
squared). The $\Delta_{\sigma^k}S_{\sigma^k}$ is
discontinuity of this metric induced from a certain pair of
the different 4-simplices sharing this $k$-simplex when
passing across $\sigma^k$ from one of these 4-simplices to
another one. Vanishing metric discontinuities on the
3-faces are conditions to be imposed on the Regge calculus
with independent 4-simplex metrics to get usual Regge
calculus. The $\delta$-functions of discontinuities on the
2-faces and 1-faces (links) serves to cancel effect of the
cycles enclosing the triangles and leading to the
singularities of the type of $\delta$-function squared.
Occurrence of the $\delta$-function in the denominator
means that the same function is contained in the numerator
and is thereby cancelled.

The $\delta_{\rm cont}$ imposes the constraints required
to "glue" together different 4-simplex metrics. The
constraints on the scalar areas have been discussed in
\cite{Mak,MakWil}. In our case of area tensors a new
possibility arises to get a system of bilinear constraints.
This possibility is just connected with used by us
extension of the set of area tensors to the frames of all
the 4-simplices containing a given triangle. Let $\bv_1$,
$\bv_2$, $\bv_3$ and $\bv^\prime_1$, $\bv^\prime_2$,
$\bv^\prime_3$ be triads of area vectors (3-vector
projections of selfdual parts of area tensors) in the
3-face induced from the two 4-simplices sharing this face.
Upon expressing the lengths in terms of area tensors the
corresponding $\delta$-factor in $\delta_{\rm cont}$ can be
rewritten as
\begin{equation}                                        
\label{delta-vv}
[\bv_1\times\bv_2\cdot\bv_3]^4\delta^6(\bv_{\alpha}\cdot\bv
_{\beta} - \bv^{\prime}_{\alpha}\cdot\bv^{\prime}_{\beta}).
\end{equation}

\noindent The overall set of constraints defined by $\delta
_{\rm metric}$ and $\delta_{\rm cont}$ turns out to be
bilinear w. r. t. the area tensors. These constraints
considered in our work \cite{Kha3} single out Regge
calculus hypersurface in the space of all arbitrary sets of
area tensors, $\{v\}$.

The role of $\delta_{\rm cont}$ in our estimate is,
roughly, in equating the scales of area tensors of the leaf
and diagonal triangles $x_1$ and $x_2$ in each pair of the
neighbouring 4-simplices. Besides $x$, we have the scale of
area tensors of the $t$-like triangles $\varepsilon$, and
it is important that at $\epsilon$ $\ll$ $x$ the
$\delta_{\rm cont}$ is invariant w. r. t. rescaling
$\varepsilon$ and, separately, overall rescaling the scales
$x$ in the different 4-simplices. To show this, analyse the
factor (\ref{delta-vv}) for the 3-faces of the different
types, $t$-like and leaf/diagonal ones.

\begin{figure}
\label{face}
\begin{picture}(200,200)(0,20)
\put (150,60){\line(1,1){120}}
\put (150,60){\line(2,1){160}}
\put (150,60){\line(3,1){120}}
\put (150,60){\line(1,0){160}}
\put (150,60){\line(2,-1){80}}
\put (270,100){\line(0,1){80}}
\put (270,100){\line(1,-1){40}}
\put (270,180){\line(1,-1){40}}
\put (270,180){\line(1,-3){40}}
\put (310,60){\line(0,1){80}}
\put (230,20){\line(1,4){40}}
\put (230,20){\line(1,2){40}}
\put (230,20){\line(2,3){80}}
\put (230,20){\line(2,1){80}}
\put (310,135){$~k^{+}$}
\put (270,180){$~i^{+}$}
\put (139,52){$m$}
\put (270,100){$~i$}
\put (308,55){$~k$}
\put (230,9){$l$}
\end{picture}
\caption{Diagonal 3-face $(i^{+}klm)$,
common for the 4-simplices $(ii^{+}klm)$ and $(i^{+}kk^{+}lm)$.}
\end{figure}

If the 3-face is leaf or diagonal one then some of
$\bv_{\alpha}$, $\bv^{\prime}_{\alpha}$ involved in
(\ref{delta-vv}) may turn out to be the algebraic sums of
those $\bv_{\sigma^2|\sigma^4}$ which are taken as
independent field variables, as mentioned in the footnote
after equation (\ref{tetrad}). For our regular way of
constructing the full 4D Regge manifold from the 3D leaves
a 4-simplex has the general form $(ii^+ABC)$ with $t$-like
edge $(ii^+)$ and each of other three vertices $A$, $B$,
$C$ laying either in the current $t$ leaf ($k$, $l$, $m$,
\dots ) or in the next-in-$t$ leaf ($k^+$, $l^+$, $m^+$,
\dots ). Take as independent area tensors in the 4-simplex
those ones of the 6 triangles containing, e. g., the vertex
$i$; then others (which are future in $t$) are expressed on
using closure relations. Let the 3-face $(i^+klm)$ be
shared by the 4-simplices $(ii^+klm)$ and $(i^+kk^+lm)$.
Then, in $\delta_{\rm cont}$, we need to compare future in
$t$ area tensors in the former 4-simplex (which are sums of
independent tensors $\pi$ and $\tau$ in this 4-simplex)
with (some combinations of) independent tensors $\pi$ in
the latter 4-simplex. In more detail, we substitute
\begin{eqnarray}                                        
\bv_1 = \bpi_{(i^+mk)il} = \bpi_{(imk)i^+l} +
\btau_{(ii^+m)kl} - \btau_{(ii^+k)lm}, & & \bv^{\prime}_1 =
\bpi_{(i^+mk)k^+l},\nonumber\\
\bv_2 = \bpi_{(i^+kl)im} = \bpi_{(ikl)i^+m} +
\btau_{(ii^+k)lm} - \btau_{(ii^+l)mk}, & & \bv^{\prime}_2 =
\bpi_{(i^+kl)k^+m},\\
\bv_3 = \bpi_{(i^+lm)ik} = \bpi_{(ilm)i^+k} +
\btau_{(ii^+l)mk} - \btau_{(ii^+m)kl}, & & \bv^{\prime}_3 =
\bpi_{(i^+lm)kk^+} = \nonumber\\
& & \hspace{-12mm}\bpi_{(klm)i^+k^+} - \bpi_{(i^+kl)k^+m} -
\bpi_{(i^+mk)k^+l}\nonumber
\end{eqnarray}

\noindent in terms of independent variables in this
example. Certain sign conventions are implied in the sums.
If $\varepsilon$ $\ll$ $x$ then $\btau$ can be neglected as
compared to $\bpi$ in these sums, and the two scales
decouple.

If the 3-face across which metric discontinuity in
(\ref{delta-vv}) is taken is $t$-like, independent area
vectors for this face are those for two $t$-like $\btau$
and one leaf/diagonal $\bpi$ 2-faces. For example, for
$\sigma^3$ = $(ii^+kl)$ shared by the 4-simplices
$(ii^+klm)$ and $(ii^+kln)$ we take
\begin{eqnarray}                                        
\bv_1 = \bpi_{(ikl)i^+m}, & & \bv^{\prime}_1 =
\bpi_{(ikl)i^+n},\nonumber\\
\bv_2 = \btau_{(ii^+k)lm}, & & \bv^{\prime}_2 =
\btau_{(ii^+k)ln},\\
\bv_3 = \btau_{(ii^+l)km}, & & \bv^{\prime}_3 =
\btau_{(ii^+l)kn}.\nonumber
\end{eqnarray}

In both cases, those ones of $t$-like or leaf/diagonal
3-face, corresponding factor in $\delta_{\rm cont}$ turns
out to be invariant w. r. t. the overall rescaling  $x$ in
the different 4-simplices. In our one-dimensional model of
estimating this corresponds to factor $x_1\delta (x_1-x_2)$
for the scales $x_1$, $x_2$ on the two neighbouring
4-simplices. Therefore the two measures $f(x_1){\rm d}x_1$
and $f(x_2){\rm d}x_2$ are "glued" together to give
\begin{equation}                                        
xf_1(x)f_2(x){\rm d}x
\end{equation}

\noindent for the overall scale $x$.

On the whole, in our one-dimensional model of estimating we
get the measure $x^{\eta -3}{\rm d}x$ per one of
$N^{(4)}_4$ 4-simplices and the factor $x^{-2}e^{-x}$ per
one of $L^{(4)}_2$ leaf/diagonal simplices; glueing the
measures with the help of $\delta_{\rm cont}$ as above we
get
\begin{equation}                                        
e^{-L^{(4)}_2x}x^{(\eta -2)N^{(4)}_4-2L^{(4)}_2-1}{\rm d}x.
\end{equation}

\noindent For the regular way of constructing the 4D Regge
calculus from the 3D leaves $N^{(4)}_4$ = $4N^{(3)}T$,
$L^{(4)}_2$ = $3N^{(3)}_2T$ where $T$ is the number of the
leaves. Besides that, $N^{(3)}_2$ = $2N^{(3)}_3$. This
results in the finite nonzero expectation values for $x$ at
$\eta$ $>$ 5,
\begin{equation}                                        
<x^j> = \left [{2\over 3}(\eta -5)\right ]^j
\end{equation}

\noindent (at $N^{(3)}_3T$ $\gg$ $j$). In particular, the
length scale $<\sqrt{x}>$ $\sim$ $\sqrt{\eta -5}$. At
$\eta$ $\leq$ 5 we find $<x^j>$ = 0. That is, the
functional integral is saturated by infinitely small $x$ at
such $\eta$. In other words, the system becomes continuum
dynamically. We can say that there is transition at $\eta$
= 5 between the discrete and continuum phases.

Consider the system in the limit of arbitrarily small angle
defects when the corresponding edge lengths vary slowly
from vertex to vertex (if Regge manifold possesses regular
structure). This situation also takes place if we view the
Regge manifold as some triangulation of a certain fixed
smooth Riemannian manifold and tend a typical triangulation
length to zero, i. e. in the continuum limit. It has been
proven \cite{Kha4} that modulo partial use of the equations
of motion the area tensor Regge calculus results in the
continuum limit in the area-generalised Hilbert-Palatini
form of GR such that upon postulating the tetrad form of
area tensors we get usual GR. In the functional integral
integrations over ${\cal D}\Omega$ reduce essentially to
${\rm d}^{24}\omega^{ab}_{\lambda}$ per point, $\omega^{ab}
_{\lambda}$ being certain combinations of generators of
$\Omega$. This integration gives $(\det{\|g_{\lambda\mu}
\|})^{-3}$ {\it per vertex}. This should be combined with
$(\det{\|g_{\lambda\mu}\|})^{(\eta -7)/2}{\rm d}^{10}
g_{\lambda\mu}$ {\it per 4-simplex} which follows upon
integrating out $\delta_{\rm metric}$. From dimensionality
arguments and invariance properties of $\delta_{\rm cont}$
w. r. t. the rescaling $g_{\lambda\mu}$ combining separate
measures of such form yields
\begin{equation}                                        
(\det{\|g_{\lambda\mu}\|})^{\alpha}{\rm d}^{10}
g_{\lambda\mu},~~~~\alpha = \left ({\eta\over 2} - 1\right
)n - {11\over 2}.
\end{equation}

\noindent Here $n$ = $N^{(4)}_4/N^{(4)}_0$, the number of
the 4-simplices per point. Thus $\alpha$ is not an
universal constant for the given theory with certain
$\eta$. Only if we restrict ourselves to summation in the
path integral over Regge manifolds with similar uniform
structure, we get a definite $\alpha$. E. g., for the
manifold with fixed simplest periodic structure
\cite{RocWil} consisting topologically of hypercubes each
composed of 24 4-simplices $n$ = 24 which gives $\alpha$
$>$ $61/2$ in the discrete phase $\eta$ $>$ 5 (though,
general combinatorial estimate gives $n$ $>$ $5/2$ which
results in weeker bound $\alpha$ $>$ $-7/4$ at $\eta$ $>$
5). In the real case of summation over all the structures
we'll find different effective $\alpha$ for the different
functionals averaged since the largest contribution will be
provided by different structures for the different
functionals. Thus, the theory stable at small edge lengths
corresponds at large distances in average to rather large
$\alpha$.

To summarize, we have discussed the quantum measure based
on the assumption that Regge calculus is a kind of the
state of some extended area tensor system with the known
quantum measure ${\rm d}\mu_{\rm area\, extended}$. The
measure of interest reads
\begin{equation}                                        
{\rm d}\mu_{\rm Regge} = \delta_{\rm cont}
\delta_{\rm metric}{\rm d}\mu_{\rm area\, extended}.
\end{equation}

\noindent Here $\delta_{\rm cont}\delta_{\rm metric}$ can
be fixed uniquely up to a parameter $\eta$ in $\delta_{\rm
metric}$. For sufficiently large $\eta$ edge length
expectation values are nonzero and finite (of the order of
Plank length). In the case of the fixed periodic structure
of the Regge manifold the theory at large distances looks
as GR with quantum measure $(\det{\|g_{\lambda\mu}\|})
^{\alpha}{\rm d}^{10}g_{\lambda\mu}$ at rather large
$\alpha$, although $\alpha$ depends on this structure and
is not an universal constant.

\bigskip

The present work was supported in part by the Russian
Foundation for Basic Research through Grant No.
03-02-17612.

\end{document}